\documentclass[twocolumn,preprintnumbers,amsmath,amssymb]{revtex4}
\usepackage{color}
\usepackage{graphics}
\usepackage{dcolumn}
\usepackage{bm}

\renewcommand\textfraction 0
\renewcommand\topfraction 1
\renewcommand\bottomfraction 1
\begin{document}
\title{Origin of spin-glass behavior of Zn$_{1-x}$Mn$_x$O} \author{S. Kolesnik, B. Dabrowski,  and J.
Mais } \affiliation{Department of Physics, Northern Illinois University, DeKalb, IL 60115}
\date{\today}
\begin{abstract}
ac susceptibility has been studied for polycrystalline Zn$_{1-x}$Mn$_x$O. Stoichiometric samples
demonstrate Curie-Weiss behavior, which indicates mostly antiferromagnetic interactions. Magnetic
susceptibility can be described by a diluted Heisenberg magnet model developed for semimagnetic
semiconductors. High-pressure oxygen annealing induces spin-glass like behavior in
Zn$_{1-x}$Mn$_x$O by precipitation of ZnMnO$_3$ in the paramagnetic matrix.

\vspace*{.5cm} Keywords: Diluted magnetic semiconductor, Zn$_{1-x}$Mn$_x$O, synthesis, spin-glass
behavior\\
\end{abstract}


\maketitle

The incorporation of spin into semiconductor electronics requires fabrication of ferromagnetic
semiconductors with the Curie temperature higher than room temperature. Theoretical predictions of
room temperature ferromagnetism in diluted magnetic semiconductors \cite{Dietl00} recently brought
wide attention to this class of materials. According to these calculations, $p$-type
Zn$_{1-x}$Mn$_x$O is a promising candidate for a room temperature ferromagnet. Ab initio band
calculations \cite{Sato00a} predict ferromagnetism to be stable in $p$-type Zn$_{1-x}$Mn$_x$O, and
antiferromagnetism in $n$-type Zn$_{1-x}$Mn$_x$O. On the other hand, a ferromagnetic phase has
been predicted for $n$-type ZnO substituted with Fe, Co, or Ni. \cite{Sato00b} Various
substitutions (B, Al, Ga, In, Si, and F) in the parent compound ZnO can increase its natural
$n$-type conduction, caused by oxygen vacancies and Zn interstitials. \cite{Minegishi97} However,
only a few attempts to introduce $p$-type conduction into ZnO have been so far successful. This
includes nitrogen doping \cite{Minegishi97} and, the theoretically proposed \cite{Yamamoto99} Ga
and N codoping. \cite{Joseph99} Preparation of $p$-type ZnO could find application in
optoelectronic devices (e. g. solar cells, short-wavelength light emitting diodes) and also allow
fabrication of transparent $p-n$ junctions.

Pulsed-laser deposited Zn$_{1-x}$Mn$_x$O thin films without intentional carrier doping show
spin-glass behavior. \cite{Fukumura01} According to this study, Mn can be dissolved in the ZnO
matrix to over 35\%. ZnO films doped with other transition metals (Co, Cr, Ni) have been reported
to be ferromagnetic in case of Co doping. \cite{Ueda01} Ferromagnetism, however, was observed for
only 10\% of studied thin films. Magnetic properties of bulk Zn$_{1-x}$Mn$_x$O have not yet been
reported.

\begin{figure}[!]
\resizebox{8.5cm}{!}{\includegraphics{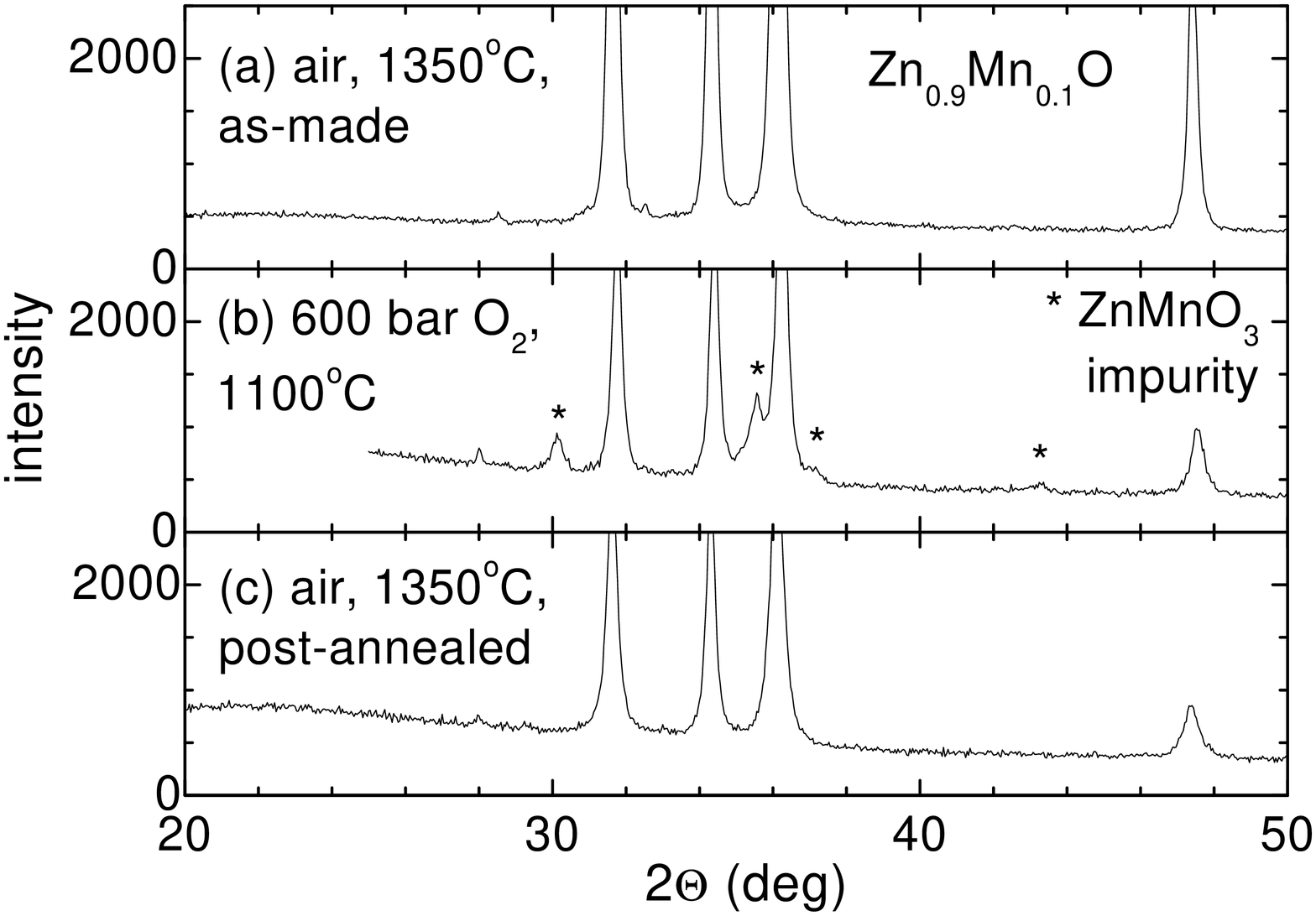}}
 \caption{\label{xray} X-ray diffraction patterns for
Zn$_{0.9}$Mn$_{0.1}$O samples synthesized in air (a), high-pressure oxygen-annealed (b), and
subsequently annealed in air (c). }
\end{figure}

In this study we investigate polycrystalline Zn$_{1-x}$Mn$_x$O $(x = 0.05 - 0.20)$. The studied
samples were prepared in air at $T = 1350^{\circ}$C using a standard solid-state reaction
technique. The samples were then annealed in an atmosphere of different gases (Ar, H$_2$, as well
as O$_2$ under high pressure of 600 bar).  ac susceptibility and dc magnetization were measured
using a Physical Property Measurement System (Quantum Design). X-ray diffraction spectra have been
collected using a Rigaku X-ray diffractometer. The samples are mostly dark green. Annealing in a
reducing H$_2$ atmosphere changes their color to orange or brown, which points at the decrease of
the energy gap below the value of 3.2 eV characteristic for ZnO. X-ray diffraction data show that
the $x \leq 0.1$ samples annealed in air are single-phase with the wurtzite structure as of ZnO.
Zn$_{1-x}$Mn$_x$O with $x = 0.15$ and 0.2 show small amount of spinel-like ZnMn$_2$O$_4$. This
observation suggests that the solubility limit in air is reached near $x = 0.1$, significantly
lower than reported for Zn$_{1-x}$Mn$_x$O thin films. \cite{Fukumura01} High-pressure oxygen
annealing causes appearance of the ZnMnO$_3$ second phase that can be removed by subsequent
annealing in air. Fig. \ref{xray} shows X-ray diffraction patterns for the Zn$_{0.9}$Mn$_{0.1}$O
sample for such a sequence of annealings. We observe the presence of impurity peaks for the
high-pressure annealed sample.

\begin{figure}[!]
\resizebox{8.5cm}{!}{\includegraphics{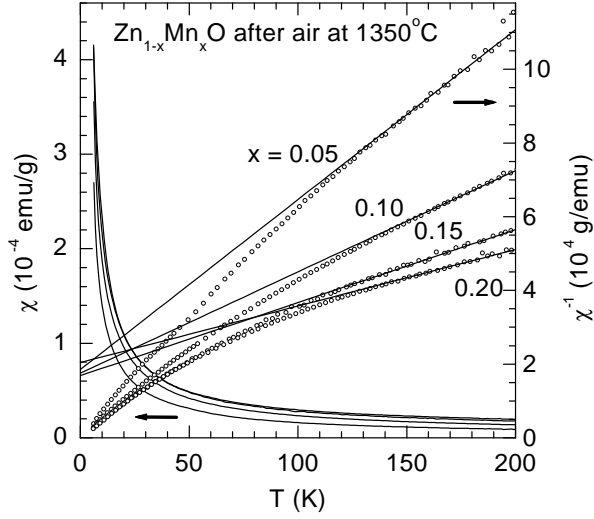}} \caption{\label{ZMair} ac susceptibility (curved
continuous lines) and inverse ac susceptibility (open symbols) for Zn$_{1-x}$Mn$_x$O samples
synthesized in air. Straight lines present results of the linear fit to the temperature dependence
of the inverse susceptibility. }
\end{figure}

Magnetic susceptibility for Zn$_{1-x}$Mn$_x$O is presented in Fig. \ref{ZMair}.
ac susceptibility of the Mn-doped ZnO at temperatures $T = 120 - 350$ K resembles Curie-Weiss
behavior that is also characteristic for other Mn-containing semimagnetic semiconductors.
\cite{Spalek86}  In Fig. \ref{ZMair}, we also present inverse ac susceptibility. In the
temperature range 150 - 300 K, we have used a linear fit to  our inverse susceptibility data.
 This linear fit, when extrapolated down to lower temperatures, intersects the $\chi^{-1} = 0$ axis at a negative temperature. This result
indicates the presence of antiferromagnetic interactions in the Zn$_{1-x}$Mn$_x$O samples.
 At lower temperatures, inverse ac susceptibility curves toward a temperature close to
zero. This points to development of additional antiferromagnetic interactions between next
neighbor Mn ions, which may turn on at lower temperatures. Similar nonlinear behavior of inverse
susceptibility at low temperatures was reported in other semimagnetic semiconductors containing
Mn. \cite{Spalek86}. The annealing in Ar or Ar/H$_2$ mixture under atmospheric pressure does not
significantly change the value of ac susceptibility.
 \begin{table*}[htb] \caption{Comparison of material parameters determined from ac
susceptibility for Zn$_{1-x}$Mn$_x$O with other semimagnetic semiconductors containing Mn.}
\begin{tabular}{cccccc}
\\
\hline \hline
Material & $\Theta_0$ (K) & $C_M$ (emu K/mol) & $2J_1/k_B$ (K) & spin $S$ & Source\\
\hline
Cd$_{1-x}$Mn$_x$Se & -743$\pm$15 & 4.81$\pm$0.1 & 21.2$\pm$0.4 & 2.64$\pm$0.06 & Ref. \cite{Spalek86}\\
Cd$_{1-x}$Mn$_x$Te & -470$\pm$34 & 4.27$\pm$0.17 & 13.8$\pm$0.3 & 2.47$\pm$0.05 & Ref. \cite{Spalek86}\\
Zn$_{1-x}$Mn$_x$O &  -961$\pm$49 &  4.07$\pm$0.2 &  27.5$\pm$1.4 &  2.5 & This work\\
Zn$_{0.64}$Mn$_{0.36}$O & -1900 & 7.6 & 30 & 3.4 & Ref. \cite{Fukumura01} \\
\hline \hline\\
 \end{tabular}
 \end{table*}

\begin{figure}[!]
\resizebox{12cm}{!}{\includegraphics{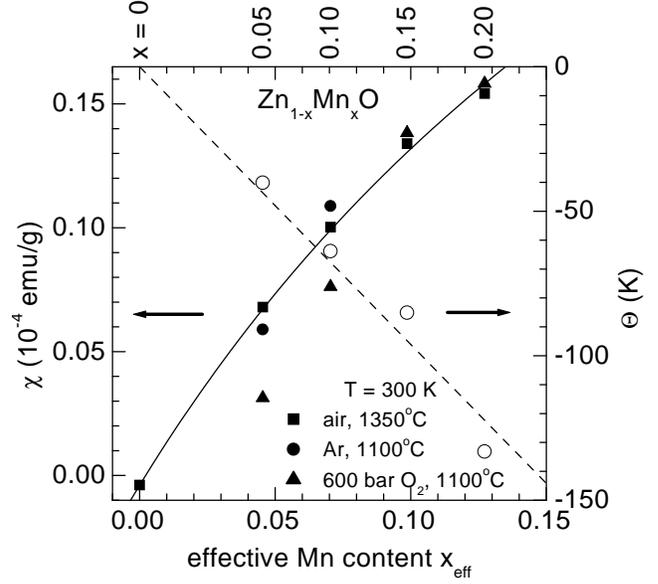}}
 \caption{\label{chix} Curie-Weiss temperatures
$\Theta$ and ac susceptibility at $T = 300$ K for Zn$_{1-x}$Mn$_x$O samples. The solid curve shows
ac susceptibility calculated according to Eqs. (1) and (2), with $S = 5/2$ and derived $\Theta_0 =
- 961$ K. The dashed line is a linear fit to the $\Theta(x_{eff})$ data. Nominal Mn content $x$ is
shown above the top axis.}
\end{figure}

 We have analyzed ac susceptibility results within the framework of the diluted Heisenberg
antiferromagnet theory of Spa{\l}ek {\em et al.} \cite{Spalek86} At higher temperatures, ac mass
susceptibility can be described by the formula

\begin{equation}
 \chi = \frac{C_0x}{T-\Theta(x)},
\end{equation}
where $C_0$ is the Curie constant defined as
\begin{equation}
 C_0 = \frac{N(g\mu_B)^2S(S+1)}{3k_B\rho},
\end{equation}
$N$ - the number of cations per unit volume, $S = 5/2$ - effective spin of Mn$^{2+}$ ion, $\rho$ -
mass density (although the lattice parameters change slightly with substitution of Mn, here we use
the density $\rho = 5.55$ g/cm$^3$ of pure ZnO), $\Theta(x) = \Theta_0 \cdot x$ - Curie-Weiss
temperature. The $\Theta_0$ constant is related to the exchange integral between the nearest Mn
neighbors $J_1$,
\begin{equation}
\frac{2J_1}{k_B} = \frac{3\Theta_0}{zS(S+1)},
\end{equation}
where $z = 12$ is the number of nearest neighbors in the wurtzite
structure of Zn$_{1-x}$Mn$_x$O.

ac susceptibility at $T = 300$ K for all studied samples is compiled in Fig. \ref{chix}.
First, pure ZnO is diamagnetic. Our measurements give us a value of ac susceptibility, which is
15-20\% greater than the handbook value of $0.33 \cdot 10^{-6}$ emu/g. \cite{Handbook00} This
value does not change after argon annealing. The Mn- doped ZnO is paramagnetic. ac susceptibility
increases with increasing Mn content $x$. Again, there is no significant influence of argon
annealing on the ac susceptibility. High-pressure oxygen annealed samples show a decrease of room
temperature susceptibility, especially large for small $x$. Subsequent annealing in air increases
ac susceptibility to the initial values. The measured ac susceptibility is significantly lower
than the theoretical expectation [Eqs. (1) and (2)] for a given nominal Mn content $x$. We suggest
that due to disorder in our polycrystalline samples, not all Mn ions participate in superexchange
interactions. Therefore, by putting $C_0$ from Eq. (2) and the measured $\chi$ and $\Theta$ into
Eq. (1) we calculated the effective Mn content $x_{eff}$ for each nominal Mn content $x$.
 In Fig. \ref{chix}, we present the values of the Curie-Weiss temperatures, obtained from the linear fit to the inverse
 susceptibility data for the air-annealed samples. From the linear  dependence of $\Theta$ vs. $x_{eff}$, and assuming that
 $\Theta(x_{eff}=0) = 0$, we derived the value of the exchange parameter $\Theta_0 = - 961\pm
 49$ K. This value is close to previously reported values of  $\Theta_0$ for other semimagnetic semiconductors, but is much less
 than $\Theta_0 = - 1900$ K, determined by Fukumura {\em et al.} \cite{Fukumura01}
The solid line in Fig. \ref{chix} represents the ac susceptibility dependence on $x_{eff}$,
according to Eq. (1), using the our derived value of $\Theta_0$.
 Better agreement between the values of nominal $x$ and effective $x_{eff}$ should be obtained for single crystals of these materials, but crystal
growth of Zn$_{1-x}$Mn$_x$O with $x \geq 0.01$ from flux is difficult.\cite{Ohashi99}

In Table I we compare the results determined from ac susceptibility for our samples with those for
several semimagnetic semiconductors. Taking Mn$^{2+}$ spin $S = 5/2$, our values of  $\Theta_0$
and $2J_1/k_B$ are similar to other Mn-containing semiconductors. This supports the observation of
the dominating role of superexchange in our materials.

\begin{figure}[!]
\resizebox{18cm}{!}{\includegraphics{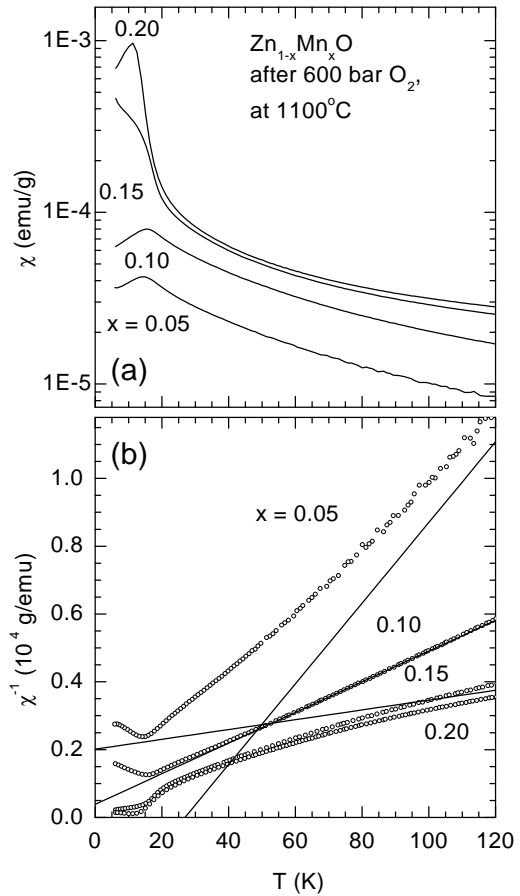}} \caption{\label{HP} (a) - ac susceptibility and (b) -
inverse ac susceptibility for Zn$_{1-x}$Mn$_x$O samples after high-pressure oxygen annealing.}
\end{figure}

 The influence of high-pressure oxygen annealing on ac susceptibility is presented in Fig.
\ref{HP} for our Zn$_{1-x}$Mn$_x$O samples.
Magnetic properties are dramatically changed after the annealing. For almost all Zn$_{1-x}$Mn$_x$O
samples, we observe a cusp in the temperature dependence of ac susceptibility [Fig. \ref{HP}(a)],
which indicates a transition to the spin-glass state. The freezing temperature of the spin-glass
is equal to 11 - 16 K, depending on the manganese content. This value is consistent with $T_f =
13$ K, noted by Fukumura {\em et al.} \cite{Fukumura01} for a Zn$_{0.64}$Mn$_{0.36}$O thin film.
For high-pressure oxygen-annealed samples with $x = 0.10$, inverse ac susceptibility is now linear
down to $T_f$ [Fig. \ref{HP}(b)]. The extrapolation of $\chi^{-1}(T)$ for this sample gives
negative Curie-Weiss temperature $\Theta$, which is close to $T_f$. Samples with $x = 0.15$ and $x
= 0.20$ show unchanged values of $\Theta$. $\Theta$ is radically changed for $x = 0.05$, becoming
positive, which indicates the presence of ferromagnetic interactions in this material.

We can correlate the observed spin-glass behavior with the presence of ZnMnO$_3$ inclusions in our
samples, detected by X-ray diffraction. The sample with $x = 0.15$ has the least amount of
ZnMnO$_3$. At the same time, no clear cusp in ac susceptibility is observed for this sample; only
a change of slope at $T_f$ can be seen. The presence of ZnMnO$_3$ impurity can be suppressed by
subsequent annealing in air. The paramagnetic character of Zn$_{1-x}$Mn$_x$O samples is restored
after this annealing in air.

\begin{figure}[!]
\resizebox{12cm}{!} {\includegraphics{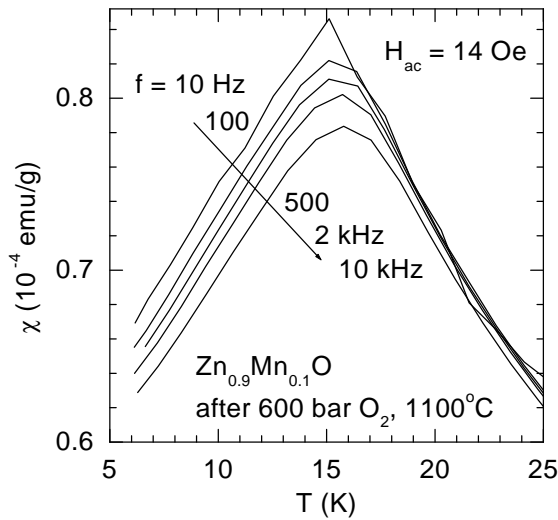}}
 \caption{\label{freq} Temperature dependence of ac
susceptibility  for Zn$_{0.9}$Mn$_{0.1}$O sample  at several frequencies.}
\end{figure}

\begin{figure}[!]
 \resizebox{12cm}{!} {\includegraphics{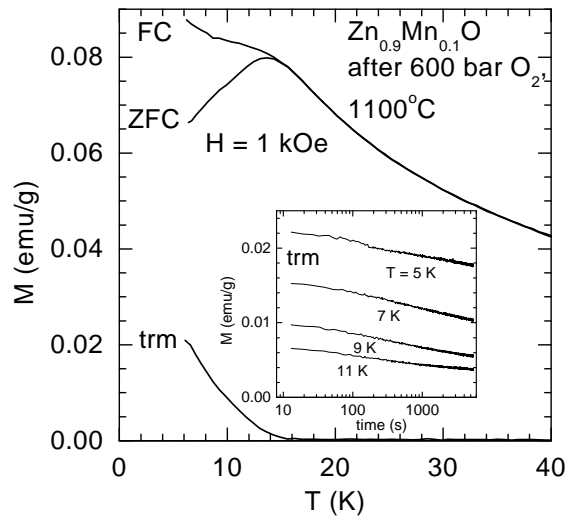} }
 \caption{\label{zfr} ``Zero-field-cooled'' (ZFC), ``field-cooled'' (FC), and thermoremanent magnetization (trm) for  Zn$_{0.9}$Mn$_{0.1}$O sample.
 Inset shows time dependence of thermoremanent magnetization at several temperatures.}
\end{figure}

 In Fig. \ref{freq},
we present ac susceptibility for a Zn$_{0.9}$Mn$_{0.1}$O sample measured at several frequencies in
an ac magnetic field of constant amplitude $H_{ac} = 14$ Oe.
One can observe a decrease of ac susceptibility below $T_f$ with increasing frequency, and a shift
of $T_f$ towards higher temperatures. This observation confirms that the observed cusp in ac
susceptibility is related to spin-glass behavior. \cite{Mydosh93} In dc magnetization, presented
in Fig. \ref{zfr},
 we can observe a difference between
``Zero-field-cooled'' (ZFC) and ``field-cooled'' (FC) magnetization, after cooling in a zero
magnetic field measured on warming and on cooling in the magnetic field, respectively.
Thermoremanent magnetization can be observed after field cooling to a temperature below $T_f$ and
switching off the magnetic field. Thermoremanent magnetization exhibits slow decay in time, which
is presented in the inset to Fig. \ref{zfr}. All these phenomena are characteristic for spin-glass
behavior.

One has to stress that the observed spin-glass behavior in our samples is different than in other
diluted magnetic II - IV semiconductors. The main difference is that in other diluted magnetic
semiconductors $T_f$ scales with the manganese content as $\ln T_f \sim \frac{n}{3}\ln x$ or $\ln
T_f \sim \alpha x^{-1/3}$. \cite{Twardowski87} For our high-pressure annealed samples,
 $T_f$ is relatively independent of $x$. For example, $T_f$ for other Mn-substituted semiconductors
 with $x = 0.2$ is usually found in the range 4 to 10 K. \cite{Twardowski87} Our air-synthesized
 Zn$_{0.8}$Mn$_{0.2}$O does not show spin-glass behavior down to $T = 2.6$ K. This observation
 again indicate the extrinsic nature of spin-glass behavior for high-pressure annealed samples.

 In
summary, we have studied magnetic properties of Mn- doped polycrystalline ZnO samples.
Stoichiometric samples demonstrate Curie-Weiss behavior at higher temperatures, with mostly
antiferromagnetic interactions. Material parameters, determined from the ac susceptibility data
show similarities between Zn$_{1-x}$Mn$_x$O and other semimagnetic semiconductors substituted with
Mn. High-pressure oxygen annealing induces spin-glass like behavior in Zn$_{1-x}$Mn$_x$O by
precipitation of ZnMnO$_3$ in the paramagnetic matrix.

This work was supported by the DARPA/ONR and the State of Illinois under HECA. The authors would
like to thank Dr. Tomasz Dietl for valuable comments.

\end{document}